\documentclass[fleqn,usenatbib]{mnras}
\usepackage[T1]{fontenc}

\DeclareRobustCommand{\VAN}[3]{#2}
\let\VANthebibliography\thebibliography
\def\thebibliography{\DeclareRobustCommand{\VAN}[3]{##3}\VANthebibliography}

\usepackage{graphicx}	% Including figure files
\usepackage{amsmath}	% Advanced maths commands
\usepackage{amssymb}	% Extra maths symbols
\usepackage{booktabs} % MCS
\usepackage{soul}   %MCS
\usepackage{newtxtext,newtxmath}

\newcommand{\Msun}{M_\odot}
\newcommand{\Mstar}{M_\star}
\newcommand{\Mhalo}{M_{\rm halo}}
\newcommand{\smf}{\Mstar/\Mhalo}
\newcommand{\abg}{$\alpha\beta\gamma$}
\newcommand{\mcs}[1]{{\textcolor{cyan}{MS: #1}}}
\newcommand{\mbk}[1]{\textcolor{red}{MBK: #1}}
\newcommand{\ag}[1]{\textcolor{green}{AG: #1}}
\newcommand{\sidm}{SIDM$_{1}$}
\newcommand{\sidma}{SIDM$_{0.1}$}
\newcommand{\sidmb}{SIDM$_{10}$}

\title[Central densities of dark matter halos]{Central densities of dark matter halos in FIRE-2 simulations of low-mass galaxies with cold dark matter and self-interacting dark matter}

\author[M. C. Straight et al.]{
\!Maria C. Straight,$^{1}$\thanks{NSF Graduate Research Fellow}\thanks{maria.straight@utexas.edu}
Michael Boylan-Kolchin,$^{1}$ James S. Bullock,$^2$ Philip F. Hopkins,$^{3}$
Xuejian Shen,$^{4}$ 
\newauthor
\,Lina Necib,$^{4}$ Alexandres Lazar,$^{2}$ Andrew S. Graus,$^{1}$ Jenna Samuel$^{1}$
\\
$^{1}$Department of Astronomy, The University of Texas at Austin, Austin, Texas 78712 USA\\
$^{2}$Department of Physics and Astronomy, University of California, Irvine, CA 92697 USA\\
$^{3}$TAPIR, Mailcode 350-17, California Institute of Technology, Pasadena, CA 91125, USA\\
$^{4}$Department of Physics \& Kavli Institute for Astrophysics and Space Research, Massachusetts Institute of Technology, Cambridge, MA 02139, USA\\
}

\date{Accepted XXX. Received YYY; in original form ZZZ}

\pubyear{2025}

\begin{document}
\label{firstpage}
\pagerange{\pageref{firstpage}--\pageref{lastpage}}
\maketitle

\begin{abstract}
We investigate the central density structure of dark matter halos in cold dark matter (CDM) and self-interacting dark matter (SIDM) models using simulations that are part of the Feedback In Realistic Environments (FIRE) project. For simulated halos of dwarf galaxy scale ($M_{\rm halo}(z=0)\approx 10^{10}\,M_\odot$), we study the central structure in both dissipationless simulations and simulations with full FIRE-2 galaxy formation physics. As has been demonstrated extensively in recent years, both baryonic feedback and self-interactions can convert central cusps into cores, with the former process doing so in a manner that depends sensitively on stellar mass at fixed $M_{\rm halo}$. Whether the two processes (baryonic feedback and self-interactions) are distinguishable, however, remains an open question. 
Here we demonstrate that, compared to feedback-induced cores, SIDM-induced cores transition more quickly from the central region of constant density to the falling density at larger radial scales. 
This result holds true even when including identical galaxy formation modeling in SIDM simulations as is used in CDM simulations, since self-interactions dominate over galaxy formation physics in establishing the central structure of SIDM halos in this mass regime. The change in density profile slope as a function of radius therefore holds the potential to discriminate between self-interactions and galaxy formation physics as the driver of core formation in dwarf galaxies. 
%This is a simple template for authors to write new MNRAS papers.The abstract should briefly describe the aims, methods, and main results of the paper.It should be a single paragraph not more than 250 words (200 words for Letters).No references should appear in the abstract.
\end{abstract}

\begin{keywords}
galaxies: dwarf -- galaxies: structure -- dark matter
\end{keywords}

%%%%%%%%%%%%%%%%%%%%%%%%%%%%%%%%%%%%%%%%%%%%%%%%%%

%%%%%%%%%%%%%%%%% BODY OF PAPER %%%%%%%%%%%%%%%%%%

\section{Introduction}
%This is a simple template for authors to write new MNRAS papers. See \texttt{mnras\_sample.tex} for a more complex example, and \texttt{mnras\_guide.tex} for a full user guide.
%All papers should start with an Introduction section, which sets the work in context, cites relevant earlier studies in the field by \citet{Fournier1901}, and describes the problem the authors aim to solve \citep[e.g.][]{vanDijk1902}. Multiple citations can be joined in a simple way like \citet{deLaguarde1903, delaGuarde1904}.

Decades of theoretical work and increasingly precise observations have established the dark energy plus cold dark matter ($\Lambda$CDM) model as the standard cosmological paradigm, capable of explaining the large-scale structure and evolution of the universe. Yet despite the successful predictions on large scales, a number of $\Lambda$CDM predictions disagree with observations on small scales, particularly for low-mass galaxies with $\Mhalo\approx 10^{10}\Msun$ \citep{bullock_small-scale_2017}. As dark-matter-dominated systems, dwarf galaxies are key sites for testing $\Lambda$CDM assumptions about the properties of dark matter and testing alternative models such as self-interacting dark matter (SIDM). 
However, dwarf galaxies are also sensitive to baryonic feedback processes. Distinguishing between the ways that baryonic processes and dark matter self interactions affect halo structure is critical for making meaningful predictions testable by observations. 

$\Lambda$CDM simulations modeling only the dark matter component of these systems find discrepancies between predictions and observations for the number, spatial distribution, and internal structure of dwarf galaxy halos \citep{sales_baryonic_2022}. One particular challenge is the dark matter content in the centers of the halos. This has historically been known as the ``cusp-core’’ problem \citep{flores_observational_1994,moore_evidence_1994} because $\Lambda$CDM simulations with only dark matter predict dwarf galaxy dark matter density profiles that rise steeply at small radii to form dense ``cuspy’’ centers \citep{dubinski_structure_1991,crone_cosmological_1994,navarro_universal_1997,navarro_inner_2004,more_cold_1999,kuzio_de_naray_mass_2008}, 
while observations seemingly indicated constant-density dark matter cores in the centers of halos \citep{salucci_dark_2000,swaters_central_2003,simon_high-resolution_2005,spekkens_cuspcore_2005,blok_high-resolution_2008,oh_dark_2011}.
However, a more recent compilation of observational data found that the observed rotation curves of dwarf galaxies with similar masses ($\Mstar\gtrsim10^7 \Msun$) imply a diversity of central dark matter distributions \citep{oman_unexpected_2015}; for example, dwarf galaxies such as Draco appear to have more cuspy halos, while others such as Fornax have dark matter cores \citep[e.g.][]{read_case_2018,pascale_action_2018}. Proper inclusion of galaxy physics and changes to the properties of dark matter have both been proposed as potential explanations for the observed diversity in dwarf galaxy rotation curves.

Although dark matter structure is dominated by gravity on cosmological scales, galaxy formation and other baryonic processes become relevant on small scales. Dwarf galaxies are particularly sensitive to baryonic feedback processes due to their low mass ($\Mstar < 10^9 M_\odot$). Stellar feedback in dwarf galaxies pushes gas out of the galaxy in energetic outflows which cause strong fluctuations in the gravitational potential. For non-adiabatic potential fluctuations and burst timescales shorter than the local dynamical time, the impulsive gas mass loss will inject energy into the orbits of the stars and dark matter, causing the dwarf galaxy to expand and puff out in an effect known as dark matter heating \citep{collins_observational_2022}. Feedback-induced core formation is most effective with multiple impulsive mass losses, and over an extended period, bursty star formation can reduce the central densities of dark matter halos and turn cusps into cores (\citealt{read_mass_2005, mashchenko_stellar_2008,governato_cuspy_2012, pontzen_how_2012}).

However, the effects of baryonic physics also depend on the stellar mass of the galaxy. For low stellar mass fractions $\Mstar/\Mhalo \lesssim 10^{-4}$ (typically $\Mstar \lesssim 10^6 \Msun$), the dark matter halos match the cuspy NFW profiles of dark-matter-only simulations. Dark matter density profiles affected by baryonic processes only become more cored at higher stellar mass fractions, with peak core formation occurring in systems with $\Mstar/\Mhalo\approx 3-5\times 10^{-3}$ \citep{di_cintio_dependence_2014}; these halos are better described by dark matter profiles that have slopes that flatten at small radii \citep{lazar_dark_2020}.

The inclusion of full galaxy physics can resolve several challenges faced by dark-matter-only $\Lambda$CDM simulations, but some tensions still remain; see \citet{sales_baryonic_2022} for a full discussion of baryonic physics in cosmological models of dwarf galaxies. Even in simulations with baryons, reproducing a variety of rotation curves in a galaxy population remains a challenge for the $\Lambda$CDM model \citep{kuzio_de_naray_baryons_2011, oman_unexpected_2015, relatores_dark_2019, santos-santos_baryonic_2020,roper_diversity_2023}, although active galactic nuclei (AGN) may play a role in diversifying dwarf galaxy density profiles \citep{koudmani_diverse_2024}. 

Beyond baryonic solutions, modifications to $\Lambda$CDM such as self-interacting dark matter (SIDM) may resolve small-scale challenges while maintaining the large-scale success of the standard cosmological model. Elastic collisions between SIDM particles transfer heat to the inner region of the halo to create central thermalized cores \citep[e.g.][]{spergel_observational_2000,kaplinghat_dark_2016, tulin_dark_2018}. For an interaction cross section per unit mass $\sigma/m \sim 1\ {\rm cm}^2\ {\rm g}^{-1}$, this energy exchange reduces the central densities of dark matter halos to form central constant-density dark matter cores \citep{vogelsberger_subhaloes_2012,peter_cosmological_2013,fry_all_2015,elbert_core_2015}. Higher cross sections result in highly efficient heat transfer, causing gravothermal collapse \citep{kochanek_quantitative_2000,balberg_self-interacting_2002,koda_gravothermal_2011}, and resulting in a greater diversity of SIDM halo density profiles \citep{zeng_core-collapse_2022,yang_strong_2023,roberts_gravothermal_2024}. 
In analytic models and N-body simulations, SIDM has been able to explain the diverse rotation curves of galaxies
\citep{kamada_self-interacting_2017,creasey_spreading_2017,kaplinghat_dark_2020,correa_tangosidm_2022}
including the extreme cases of both Draco and Fornax \citep{sameie_self-interacting_2020}. 
For a review that discusses observational constraints on SIDM, see \citet{adhikari_astrophysical_2022}.

While both feedback and SIDM may improve agreement with observations \citep{zentner_critical_2022}, it can be difficult to distinguish between the effects of dark matter self-interactions and the effects of baryonic feedback. For simulations of Milky Way (MW)-mass galaxies, SIDM density profiles are sensitive to both baryonic concentration and self-interaction cross section \citep{sameie_impact_2018}. Recent cosmological simulations of MW-mass galaxies from the FIRE-2 project found that SIDM in simulations with baryons does not change the inner structure of MW-mass galaxies as much as SIDM-only models predicted \citep{vargya_shapes_2022}. For low-mass galaxies, \citet{vogelsberger_dwarf_2014} and \citet{fry_all_2015} found no appreciable difference between CDM and SIDM predictions in simulations including baryonic physics of dwarf galaxies with $\Mstar\approx 10^8 \Msun$. However, this stellar mass range has been identified as the mass range at which baryonic core formation is maximally efficient in dwarf galaxies \citep{di_cintio_dependence_2014, chan_impact_2015, tollet_nihao_2016}, potentially obscuring which mechanism is driving core formation. \citet{robles_sidm_2017} show that the central densities of SIDM halos in simulations with or without \mbox{FIRE-2} baryonic physics are similar, indicating that SIDM central densities may be more robust to the inclusion of baryonic physics than CDM in low-mass galaxies. Distinguishing between the effects of baryons and self-interactions requires simulating below the stellar mass range at which baryonic feedback significantly affects the halo structure. Indeed, \citet{robles_sidm_2017} find that in four simulated galaxies less massive than $\Mstar \lesssim 3\times 10^6 \Msun$, SIDM forms cores and CDM forms cusps. 

In this work, we seek to distinguish between the predictions of two potential solutions to the central density problem. Due to their shallow potential wells, dwarf galaxies are particularly sensitive to both baryonic feedback and the assumed properties of dark matter, both of which affect the central densities of their dark matter halos. We consider SIDM with elastic scattering with a velocity-independent cross section per unit mass, $\sigma/m$, in simulations of halos with $5\times10^{-5} \lesssim \Mstar/\Mhalo \lesssim 10^{-3}$, which is below the stellar mass fraction range for peak feedback-induced core formation, and in some cases is lower than the stellar mass threshold needed for the baryons to significantly affect the halo structure. We compare dark matter density profiles in eight simulations that include full baryonic physics with CDM and SIDM and consider the logarithmic slopes and analytical density profile fits to determine the differences. The different shapes of the density profiles are described using different analytical density profiles, including modified NFW \citep{navarro_structure_1996} and \citet{einasto_construction_1965} profiles. 

The paper is structured as follows: in section~\ref{sec:Methods} we describe the simulations used, in section~\ref{sec:Results} we explain the best-fit analytical profiles and a comparison of the CDM and SIDM density profile slopes, and we conclude in section~\ref{sec:Conclusions}.

%\section{Methods, Observations, Simulations etc.}
% Normally the next section describes the techniques the authors used. It is frequently split into subsections, such as Section~\ref{sec:maths} below.

\section{Methods}\label{sec:Methods}
\subsection{Simulations}
\label{subsec:sims}
This analysis uses a suite of CDM simulations first presented in \citet{fitts_fire_2017} as part of the Feedback in Realistic Environments (FIRE)\footnote{\url{http://fire.northwestern.edu}} project \citep{hopkins_galaxies_2014}. A full description of these high resolution ($m_{\rm baryon}=500 \,\Msun$, $m_{\rm DM}=2500 \,\Msun$) cosmological zoom-in  \citep{onorbe_how_2014} simulations is given in \citet{fitts_fire_2017}. Isolated halos are chosen to have virial masses of $10^{10}\, \Msun (\pm 30\%)$ at $z=0$ and are required to be separated from more massive halos by at least three times the virial radius of the more massive halo and by at least five times the virial radius of the target halo. This allows the internal baryonic physics to be studied separately from environmental effects. The simulated galaxies have stellar masses $\Mstar\approx10^5-10^7\,\Msun$, consistent with rough abundance matching estimates. Each CDM simulation has an analogous self-interacting dark matter (SIDM) version with identical initial conditions and identical physics except that dark matter particles have a self-interaction cross section $\sigma/m = 1\ {\rm cm}^2\ {\rm g}^{-1}$, following the \citet{rocha_cosmological_2013} implementation of SIDM; several of these SIDM runs were first described in \citet{robles_sidm_2017}. 

Initial conditions are generated at $z=127$ with \textsc{music} \citep{hahn_multi-scale_2011} assuming a consensus $\Lambda$CDM cosmology with cosmological parameters of $h = 0.71$, $\Omega_\Lambda = 1 - \Omega_{\rm m} = 0.734$, $\Omega_{\rm b} = 0.0449$, $n_{\rm s} = 0.963$, and $\sigma_8 = 0.801$. These parameters were originally based on analysis of the seven-year WMAP data \citep{komatsu_seven-year_2011}; recent \textit{Planck} data \citep{planck2020} have resulted in slight parameter shifts that are unimportant for the results of this work. 

The simulations are evolved using \textsc{gizmo} \citep{hopkins_gizmo_2015}; simulations that include galaxy formation physics adopt the FIRE-2 model, which is described in detail in \citet{hopkins_fire-2_2018}. Briefly, gas cooling is solved with a standard implicit algorithm described in \citet{hopkins_galaxies_2014} in which heating/cooling rates are computed from $T = 10-10^{10}\ {\rm K}$ using free-free, photo-ionization/recombination, Compton, photo-electric, metal line, molecular, fine-structure, dust collisional, and cosmic ray processes. Ionization states are tabulated from \textsc{cloudy} simulations including the effects of local radiation sources and a uniform but redshift-dependent background \citep{faucher-giguere_new_2009}. The star formation method uses a sink particle approach to form star particles from gas particles that are locally self-gravitating, self-shielding, Jeans unstable, and above a minimum density of $10^3\ \mathrm{cm^{-3}}$. Once formed, star particles are treated as single stellar populations with known age, metallicity, and mass. Simulated feedback mechanisms include Type Ia and Type II supernovae, stellar winds, photo-ionization and photo-electric heating, and radiation pressure. Feedback quantities are calculated directly from standard stellar population models \citep[\textsc{starburst99};][]{leitherer_starburst99_1999} without any subsequent adjustment or fine-tuning. The simulation suite used in this analysis uses the exact FIRE physics, source code, and numerical parameters as \citet{hopkins_fire-2_2018}. Dark matter halos are identified in post-processing using the halo finder \textsc{rockstar} \citep{behroozi_rockstar_2012}, which uses an adaptive hierarchical refinement of friends-of-friends groups in six phase-space dimensions and one time dimension for robust tracking of substructure. 

Each simulation with full FIRE-2 physics (for both CDM and SIDM) also has an analogous dark-matter-only (DMO) version. The DMO simulations have identical initial conditions except with dark matter replacing the baryons, which increases individual particle masses by a factor of $(1-f_{\rm b})^{-1}$ where $f_{\rm b} \equiv \Omega_{\rm b}/\Omega_{\rm m} = 0.168$ is the cosmic baryon fraction and the value for the cosmology used in these simulations. Quoted results for the DMO simulations use $m_{\rm p}\xrightarrow{}(1-f_{\rm b})m_{\rm p}$ and therefore $\rho(r)\xrightarrow{}(1-f_{\rm b}) \rho(r)$.
%%%%%%%%%%%%%%%%%%%%%%%%%%%%%%%%%%%%%%%%%%%%%%%%%%
\begin{figure}
    \includegraphics[width=\columnwidth]{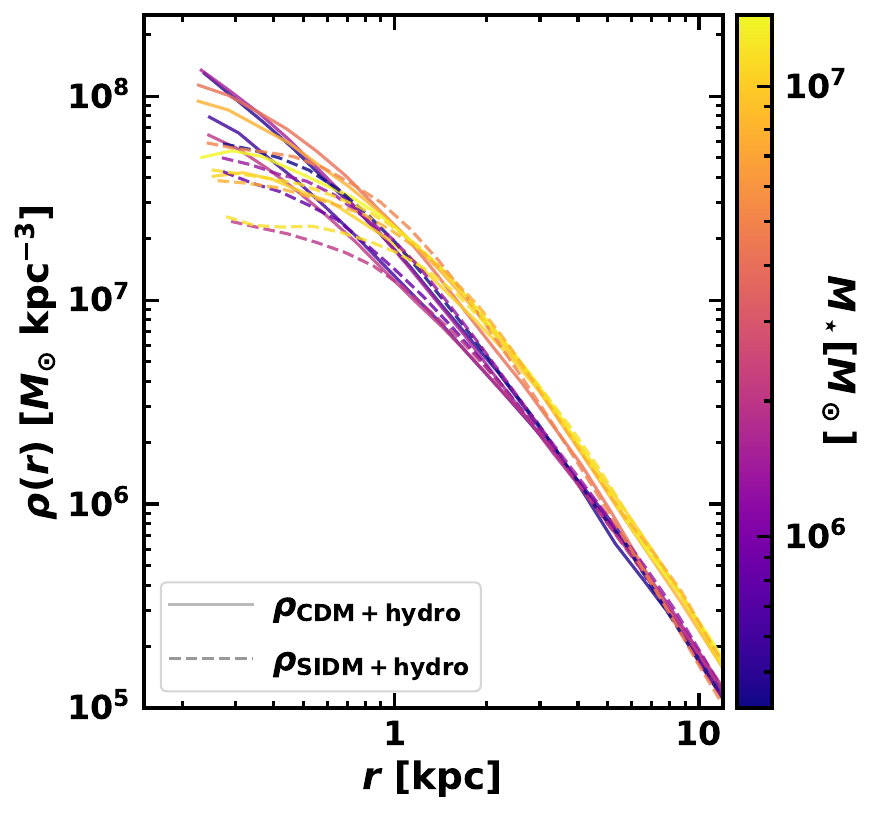}
    \caption{Dark matter halo density profiles from FIRE-2 simulations with full baryonic physics for eight galaxies of increasing stellar mass (galaxy properties in Table~\ref{tab:CDM-SIDM-parameters}). The solid lines show cold dark matter profiles (CDM+hydro) and the dashed lines show the analogous self-interacting dark matter profiles (SIDM+hydro). The profiles begin at each halo's convergence radius ($r_{\rm conv}\sim 0.2\ {\rm kpc}$) calculated  using the method in \citet{power_inner_2003}. The central regions of CDM halo density profiles have shallower slopes (more cored) in galaxies with greater stellar masses due to the increased stellar feedback. At fixed stellar mass, the SIDM halos have more cored inner densities than CDM halos.}
    \label{fig:DMhydroProfiles}
\end{figure}
%%%%%%%%%%%%%%%%%%%%%%%%%%%%%%%%%%%%%%%%%%%%%%%%%%
%%%%%%%%%%%%%%%%%%%%%%%%%%%%%%%%%%%%%%%%%%%%%%%%%%
\begin{figure}
    \centering
    \includegraphics[width=\columnwidth]{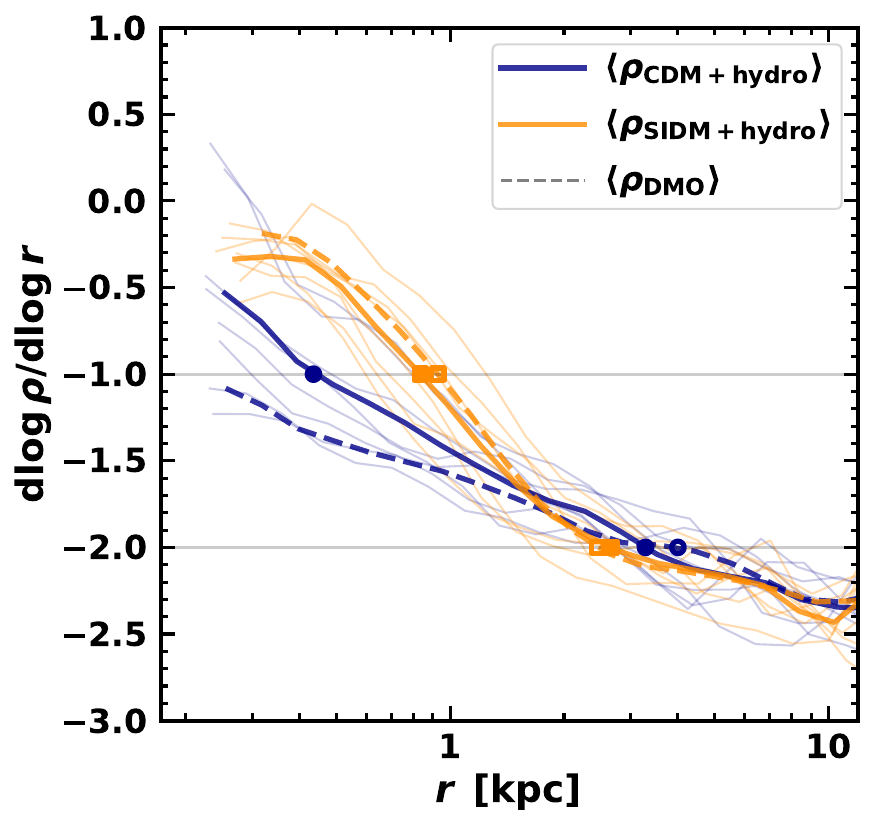}
    \caption{Individual and averaged logarithmic slopes of the density profiles for CDM (blue) and SIDM (orange) halos. Markers indicate where the logarithmic slopes of the density profiles are equal to -1 and -2. The distance between these two radii, $r_{-1}$ and $r_{-2}$, provides a measure of how quickly the profiles transition from a steep outer profile to a shallow inner core. Comparing CDM and SIDM with baryonic physics (+hydro; solid lines) and with dark matter only (DMO; dashed lines) shows that dark matter self-interactions shorten the distance between $r_{-1}$ and $r_{-2}$ more significantly than the inclusion of baryonic physics in CDM simulations. 
    }
    \label{fig:logslopes}
\end{figure}
%%%%%%%%%%%%%%%%%%%%%%%%%%%%%%%%%%%%%%%%%%%%%%%%%%

\subsection{Characterizing dark matter halos}

Dark matter halos in the simulations are defined as spherical systems with virial radius $r_{\rm vir}$ enclosing a region of average density $\Delta_{\rm vir}(z)\rho_{\rm crit}(z)$, where $\rho_{\rm crit} = 3H^2(z)/8\pi G$ is the critical density of the Universe at redshift $z$ and $\Delta_{\rm vir}(z)$ is the redshift-dependent virial overdensity defined by \citet{bryan_statistical_1998}. The halo centers are found by iteratively computing the center of mass of the dark matter particles within a sphere, re-centering at the center of mass, reducing the radius of the sphere, and computing the new center of mass until the sphere contains less than one thousand particles. As numerical relaxation affects the innermost regions of the halos, we adopt the \citet{power_inner_2003} criterion for the convergence radius $r_{\rm conv}$. Simulations that include baryons can have better or worse convergence than their DMO analogs as convergence depends on the baryonic physics rather than the baryonic particles. For the simulations we consider, $r_{\rm conv}$ is calculated using the Power criterion including both the dark matter and star particles, resulting in $\gtrsim 1000$ enclosed dark matter particles.

We construct spherically averaged density profiles of the dark matter halos using 25 logarithmically spaced bins starting from the convergence radius and extending out to the virial radius of each halo. Figure~\ref{fig:DMhydroProfiles} shows the density profiles for each of the full physics simulations considered here, with line color mapping to the stellar mass at $z=0$. We then construct profiles of the logarithmic slope of the density profile (hereafter referred to as the logslope),  
\begin{equation}
     \kappa(r)\equiv \left(\frac{\mathrm{d}\log{\rho}}{\mathrm{d}\log{r}}\right)\,,
    \label{eqn:logslope}
\end{equation}
directly from the binned density profiles. We define the radius where $\kappa$ first falls to a value of $-X$ as $r_{-X}$; we will most often be interested in $r_{-1}$ and $r_{-2}$, since $r_{-1}$ is a non-parametric proxy for the core radius, and the distance or ratio between these two radii indicates how quickly a profile transitions with increasing radius from a shallow inner core to a steep outer profile.
%%%%%%%%%%%%%%%%%%%%%%%%%%%%%%%%%%%%%%%%%%%%%%%%%%%%%%%%%%%%%%%%%%%%%%%%%%%%%%%%%%%
\begin{figure*}
    \centering
    \includegraphics[width=\columnwidth]{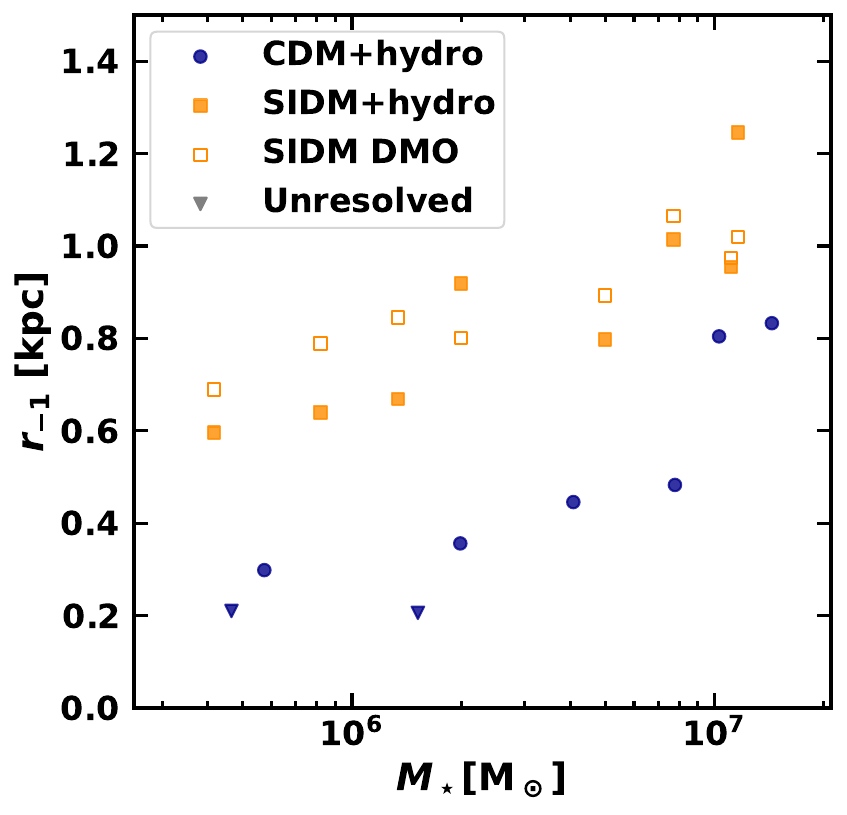}
    \includegraphics[width=\columnwidth]{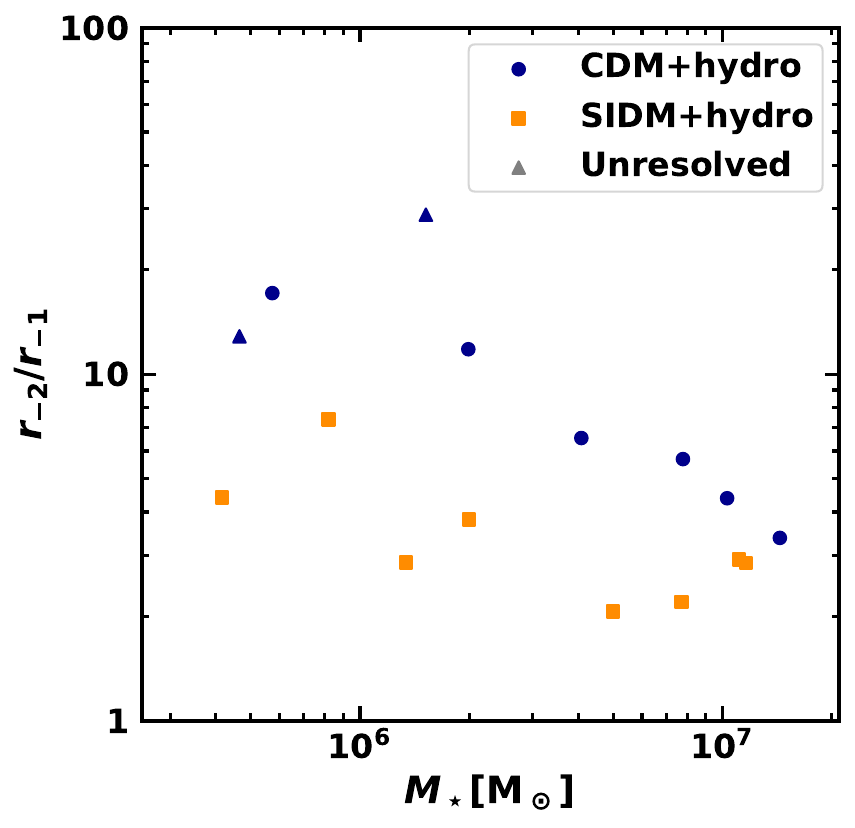}
    \caption{\textit{Left:} The radius at which the logslope of the density profile is -1, $r_{-1}$, for eight FIRE-2 classical dwarf galaxy dark matter halos simulated in CDM and SIDM with and without hydrodynamics. Both the self-interacting dark matter with baryons (SIDM+hydro; filled squares) and the dark-matter-only simulation of self-interacting dark matter (SIDM DMO; open squares, plotted at the stellar mass of the analogous SIDM+hydro runs) have higher values of $r_{-1}$ than cold dark matter (CDM+hydro; filled circles). Triangular markers indicate halos in which the density profile logslope is steeper than -1 at all radii larger than the convergence radius, characteristic of a cuspy profile. With or without the inclusion of baryons, halos of self-interacting dark matter form larger cores than halos of cold dark matter.
    \textit{Right:} The ratio between $r_{-2}$ and $r_{-1}$ for each halo. The smaller ratios for the SIDM profiles correspond to a shorter distance between the two radii, indicating a sharper transition in the density profile between the central core and outer envelope.
    }
    \label{fig:r-1}
\end{figure*}

\section{Results}\label{sec:Results}

\subsection{Logslopes in SIDM versus CDM}
Figure~\ref{fig:DMhydroProfiles} shows the density profiles for every CDM (solid) and SIDM (dashed) run with full galaxy formation physics (+hydro) considered here. The color of each line indicates the stellar mass of its central galaxy, with lower masses in purple and higher masses in yellow. The densest halos are all CDM versions, whereas the lowest density halos are all SIDM. Halos of intermediate central density can be either CDM or SIDM, but at a fixed stellar mass, the SIDM halos are clearly distinguishable from CDM versions, as they have higher densities at larger radii and transition quickly to lower densities at smaller radii. This observation motivates a consideration of the logarithmic slopes of the density profiles in the two dark matter models.  

Figure~\ref{fig:logslopes} shows the central result of this paper. Logarithmic density slopes are plotted as a function of radius for CDM (blue) and SIDM (orange) halos, both DMO (dashed) and full galaxy formation physics (solid). In each case, the thick lines show averaged results; for the full galaxy formation physics runs, we also show the results of individual profiles in thin solid lines. The CDM DMO profiles have logslopes that transition gradually with decreasing radius from isothermal to (shallow) central cusps, while the SIDM DMO profiles transition quickly with radius from an outer region that matches the CDM slope to an inner cored profile. When adding galaxy formation physics, the CDM halos show a quicker transition with radius to a shallower inner profile (solid blue), whereas the mean SIDM profile is unchanged relative to the DMO version (solid versus dashed orange lines). 

Thus, while both baryonic physics and self-interactions can lead to cores, the effects are, in principle, distinguishable: \textit{baryon-induced cores transition more slowly from a logslope of $-1$ to $-2$ compared to cores originating from self interactions.} Furthermore, the typical dark matter profile in SIDM halos is set almost entirely by self-interactions, with the physics of galaxy formation contributing (on average) very little: the dashed and solid orange lines are virtually identical in Fig.~\ref{fig:logslopes}. 

The left panel of Fig.~\ref{fig:r-1} augments the results of Fig~\ref{fig:logslopes} by showing the location of $r_{-1}$ in each galaxy. As described in \citet{fitts_fire_2017}, the stellar masses of the galaxies in our sample of halos at fixed $\Mhalo=10^{10}\,\Msun$ correlate with formation time, or alternately, with $V_{\rm max}$ or concentration: earlier forming halos have higher concentrations, higher $V_{\rm max}$ values, and therefore deeper central potentials. We see this reflected in the CDM+hydro results (blue points): a deeper gravitational potential means more star formation, which leads to more effective density reduction by feedback and a larger value of $r_{-1}$. In the case of SIDM halos, a higher initial central density leads to more self-scatterings; as a result, the halos with deeper potentials have larger core radii in the SIDM DMO runs (open orange squares). 

The left panel of Fig.~\ref{fig:r-1} also demonstrates that sizeable ($r>500\,{\rm pc}$) feedback-induced dark matter cores in the CDM runs develop only in the two galaxies with the highest stellar masses, while the other CDM profiles remain relatively cuspy. This is consistent with previous studies which have shown that feedback in the smallest galaxies does not significantly alter the cuspy dark-matter-only predictions \citep{di_cintio_dependence_2014, lazar_dark_2020}. Their analogous SIDM halos, however, predict cores for halos even in galaxies with lower stellar masses. Interestingly, the $r_{-1}$ values of the SIDM DMO runs tend to be slightly ($\sim 10-15\%$) larger than those in the SIDM+hydro runs, indicating that feedback does have a small effect on SIDM halos (and tends to soften the cores slightly). 

The right panel in Fig.~\ref{fig:r-1} quantifies how quickly the profiles in the full physics runs transition from a logslope of $-1$ to a logslope of $-2$. The SIDM runs systematically transition more quickly, with a mean value of $r_{-2} \approx 3\,r_{-1}$; for the CDM runs, we find an average value of $r_{-2} \approx 7\,r_{-1}$. Some of the most massive galaxies in SIDM have very rapid transitions, with profiles transitioning from a logslope of $-1$ to $-2$ while the radius changes by only a factor of 2. The CDM simulations exhibit a strong dependence on stellar mass, while the SIDM dependence is somewhat weaker. 

The difference in the density profile shapes between each CDM+hydro and its paired SIDM (or SIDM+hydro) run originates from the different physical processes heating the dark matter. In the case of CDM with full galaxy formation physics, \citet{pontzen_how_2012} have demonstrated that repeated episodes of impulsive energy injection from star formation feedback leads to a slow net heating of dark matter at the centers of galaxies. This effect is gradual because it requires many bursts of star formation, and it is not sharply localized within the galaxy because (1) the galaxy ``breathes'' as a result of the feedback \citep{el-badry_breathing_2016} and (2) the orbits of dark matter particles have a wide mix of eccentricities \citep{diemand2007a}, meaning the heating affects particles with a range of orbit-averaged positions.

On the other hand, the shape of SIDM cores is set by the interaction rate of the dark matter particles. The dense inner regions of the halo allow frequent interactions to keep the dark matter in thermal equilibrium, while interactions in the outer region of the halo are infrequent enough to be negligible.  Self-interacting dark matter density profile shapes are well understood from isothermal Jeans modeling: \citet{robertson2021} found their properties can be accurately captured by a model that assumes an abrupt boundary between these two regions. This boundary is defined by the radius at which all dark matter particles within have interacted at least once within the approximate age of the halo and relates to the faster core to outer halo transition seen in our simulations in SIDM relative to CDM+hydro. Comparing DMO halos of CDM and SIDM, we find that this radius approximately corresponds to the radius at which the SIDM density profile departs from the CDM profile and becomes more cored, around $r_{-1}$ for the SIDM profile.

\subsection{Analytical profiles}\label{subsec:analytical}

$\Lambda$CDM halos of all masses in simulations with only collisionless dark matter are reasonably well described by the Navarro-Frenk-White (NFW) double power law profile \citep{navarro_structure_1996}. However, the innermost regions of simulated halos deviate slightly but systematically to shallower logarithmic slopes, relative to the NFW profile \citep{navarro_inner_2004,navarro_diversity_2010}. A somewhat better fit is obtained if the logarithmic slope of the density profile $\kappa(r) \equiv \mathrm{d}\log{\rho}/\mathrm{d}\log{r}$ is assumed to vary continuously with radius: 
setting 
\begin{equation}
    \kappa(r) \equiv -2 \left(\frac{r}{r_{-2}}\right)^\alpha
\end{equation}
results in the three-parameter \citet{einasto_construction_1965} profile 
\begin{equation}
    \rho_{\mathrm{Ein}}(r) = \rho_{-2} \exp \left\{-\frac{2}{\alpha}\left[\left(\frac{r}{r_{-2}}\right)^{\alpha}-1\right]\right\}\,,
\label{eqn:Einasto}
\end{equation}
with $\rho_{-2}=\rho(r_{-2})$.
Fixing $\alpha = 0.16$ results in a good fit for typical halos in dark-matter-only (DMO) simulations \citep{navarro_inner_2004, springel2008}, resulting in an improved two-parameter analytical density profile for DMO halos relative to the NFW fit. 

To more accurately characterize halos affected by baryonic feedback, other models based on these standard dark matter profiles add a third free parameter to allow for a defined constant-density core in the innermost resolved region of the halo. For example, \citet{read_dark_2016} introduced the ``core-NFW'' profile --- where the density profile smoothly transitions from an NFW profile at large radii to a shallower profile at smaller radii, with the amount of deviation from NFW depending on the star formation history (see also \citealt{de_leo_surviving_2024} for an updated empirical calibration of the core-NFW parameter $n$). In a similar vein, and given the general better agreement between profiles in collisionless simulations and the Einasto profile (compared to NFW), \citet{lazar_dark_2020} introduced a ``core-Einasto'' profile that accounts for the effects of baryonic feedback in CDM halos:
\begin{equation}
    \rho_{\mathrm{cEin}}(r) = \tilde{\rho}_{\rm s} \exp \left\{-\frac{2}{\hat{\alpha}}\left[\left(\frac{r+r_{\rm c}}{\tilde{r}_{\rm s}}\right)^{\hat{\alpha}}-1\right]\right\}
\label{eqn:coredEinasto}
\end{equation}
where $r_{\rm c}$ is the dark matter core radius and $\tilde{r}_{\rm s}$ and $\tilde{\rho}_{\rm s}$ are radius and density free parameters. 
Setting the shape parameter $\hat{\alpha}=0.16$ \citep{gao_redshift_2008} results in a three-parameter fit. In our task of understanding which dark matter models have the potential to resolve the central density problem, this profile with its core radius parameter $r_{\rm c}$ will identify whether a dark matter halo forms a core and determine the radius of that core.
Since $\tilde{\rho}_{\rm s}$ becomes $\rho_{-2}$ as $r_{\rm c}$ approaches 0, the core-Einasto profile becomes a regular Einasto profile for halos without a core. 

Another well-studied model related to the NFW profile is the general five-parameter $\alpha\beta\gamma$-profile \citep{zhao_analytical_1996}, which takes the form
\begin{equation}
    \rho_{\alpha\beta\gamma}(r) = \frac{\rho_{\rm s}}{(r/r_{\rm s})^{\gamma_{\rm s}}\left[1+(r/r_{\rm s})^{\alpha_{\rm s}}\right]^{(\beta_{\rm s}-\gamma_{\rm s})/\alpha_{\rm s}}}
\label{eqn:alphabetagamma}
\end{equation}
where $r_{\rm s}$ and $\rho_{\rm s}$ are the scale radius and scale density, the inner and outer regions of the halo are parameterized by the logarithmic slopes $-\gamma_{\rm s}$ and $-\beta_{\rm s}$, and $\alpha_{\rm s}$ again describes the rate at which the slope changes with radius between the inner and outer regions of the halo. Fixing $\beta_{\rm s}=2.5$ and $\gamma_{\rm s} = 0$ \citep{di_cintio_mass-dependent_2014} results in a three-parameter fit with free shape parameter $\alpha_{\rm s}$. Here, we examine the agreement between our simulated density profiles and the core-Einasto profile or \abg\ profile.

\begin{figure}
    \centering
    \includegraphics[width=\columnwidth]{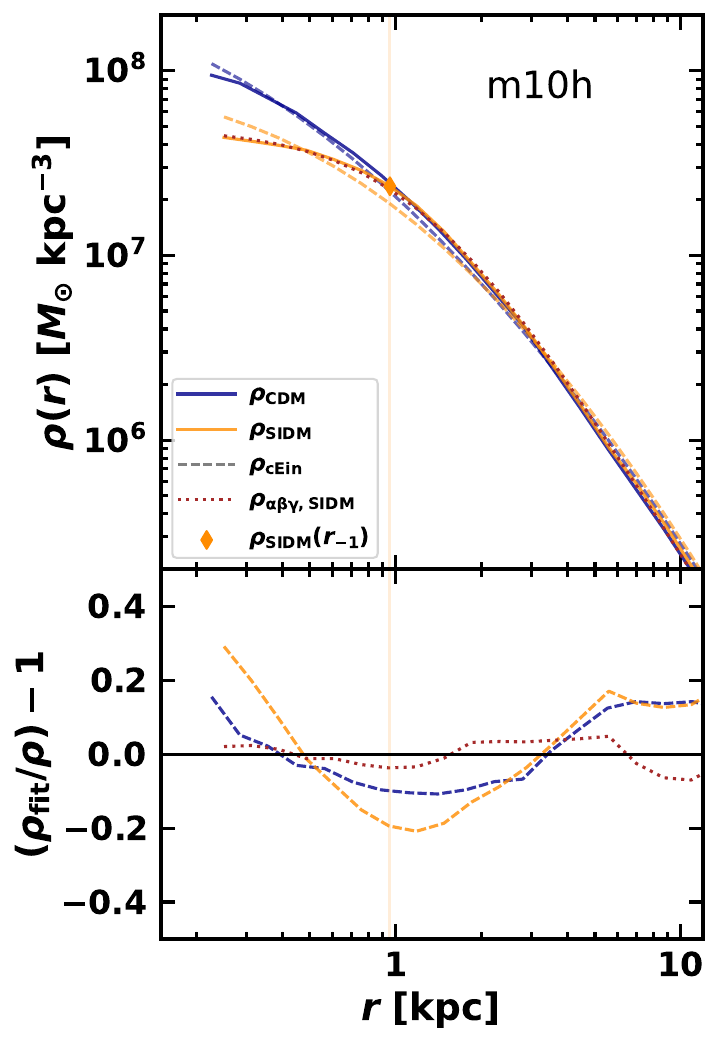}
    \caption{Density profiles and fitted analytical profiles (upper panel) and profile residuals (lower panel) for the CDM (blue) and SIDM (orange) versions of halo m10h simulated with full galaxy formation physics. The core-Einasto fit (dashed lines) does not change rapidly enough between the flatter inner region and the outer steep envelope to fit either the CDM or SIDM halos, though it performs better with the cuspier CDM halo. The residuals show that the fit is worst at $r \sim 1~{\rm kpc}$ in both cases, comparable to $r_{-1}$ (marked with a diamond and vertical line). 
    The \abg\ profile fit (dotted line) shows an improved fit compared to the core-Einasto fit to the SIDM halo's density profile. 
    The more rapid transition enabled by the $\alpha_{\rm s}$ parameter allows a good fit of the \abg\ profile to the SIDM profile at all radii.
    }
     \label{fig:cEin_vs_alpha}
\end{figure}

Figure~\ref{fig:cEin_vs_alpha} shows the CDM and SIDM density profiles of the m10h halo (solid lines) plotted with the core-Einasto fit for each profile (dashed lines) and the \abg\ fit for the SIDM profile (dotted line). The residuals shown in the lower panel of the plot show the deviations of the fits from the density profiles of the halo, revealing that the core-Einasto model fits the CDM simulation better than it fits the SIDM model. However, there is still a clear discrepancy in the fit at $\sim1\,{\rm kpc}$ in both cases. By contrast, the \abg\ profile provides an excellent fit at all radii even for the SIDM+hydro simulation, which is reflected in the smaller quality-of-fit $Q$ value (Table~\ref{tab:CDM-SIDM-parameters}).

\begin{figure}
    \centering
    \includegraphics[width=\columnwidth]{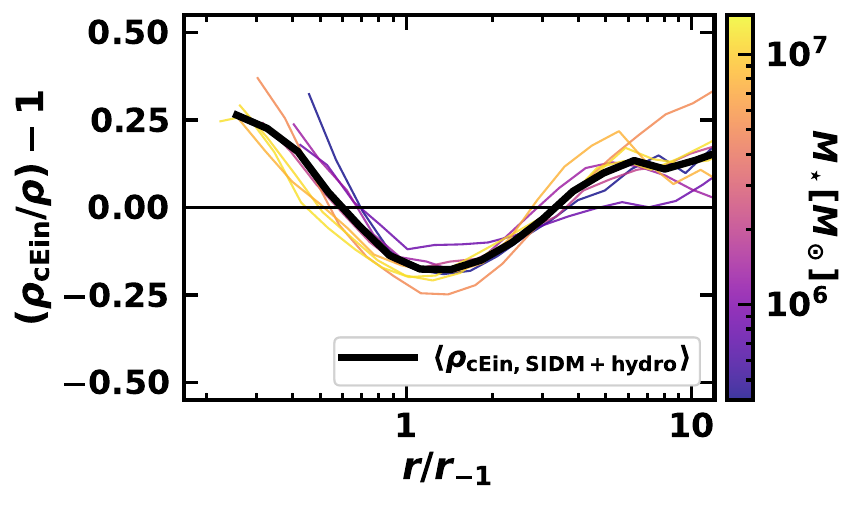}
    \caption{The core-Einasto residuals for the SIDM density profiles as a function of the radius scaled by $r_{-1}$. The residual averaged over all eight simulations is shown by the thick black line, while residuals of individual halo density profile fits are shown by lines with colors corresponding to stellar mass, as indicated by the colorbar to the right. The analytical core-Einasto profile, designed for feedback-affected halos, fails to capture the sharp transition to a core found in the density profiles of SIDM halos, particularly around $r_{-1}$, the radius at which the logslope is $-1$.}
    \label{fig:SIDM_r1}
\end{figure}

\begin{figure*}
    \centering
    \includegraphics[width=\textwidth]{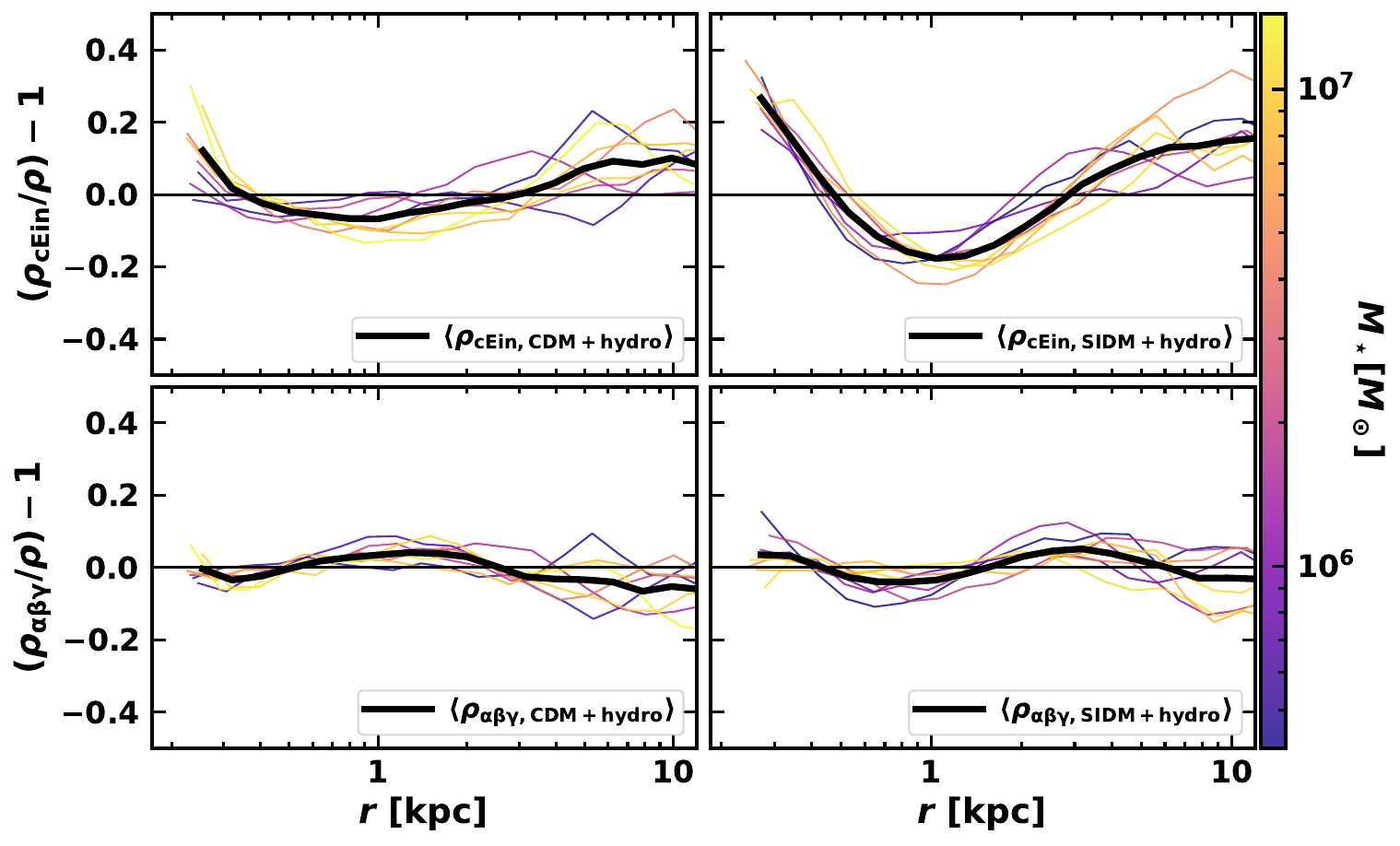}
    \caption{Profile residuals for CDM (left) and SIDM (right) for the core-Einasto profile (upper panels) and  $\alpha\beta\gamma$ profile (lower panels). For each panel, the individual profile residuals are shown as colored lines (with color corresponding to the stellar mass indicated by the colorbar on the right), while the residual averaged over all simulations is shown in thick black. The core-Einasto profile (Eq.~\ref{eqn:coredEinasto}) provides an adequate fit to our simulations' density profiles for feedback-affected CDM profiles but consistently fails to fit the shape of the corresponding SIDM profiles, with residuals of $\sim 20\%$ near $1\,{\rm kpc}$ (in the direction of the fit underpredicting the true density). The sharper turnovers of SIDM density profiles are better fit by the $\alpha\beta\gamma$ profile (Eq.~\ref{eqn:alphabetagamma}), as the $\alpha_{\rm s}$ parameter gives the freedom to more rapidly transition from the inner core to the outer envelope.
    }\label{fig:residuals}
\end{figure*}

Figure~\ref{fig:SIDM_r1} demonstrates that \textit{all} of the core-Einasto fits to the SIDM+hydro simulations of classical dwarf galaxies suffer from the same deviations: they underpredict the density near $r_{-1}$ and overpredict it at smaller radii. This common feature across all simulations in this mass regime indicates a general shortcoming of the core-Einasto model: it is unable to replicate the relatively rapid transition from the outer power-law portion to the inner density core seen in the SIDM+hydro models. The same is true, albeit to a lesser extent, even when fitting to the CDM+hydro simulations, as Fig.~\ref{fig:residuals} demonstrates. The top panels of the figure show the residuals of the core-Einasto fits to CDM+hydro (left) and SIDM+hydro (right), and in both cases, the core-Einasto fit has clear correlated deviations. The \abg\ fit, on the other hand, is able to much more accurately capture the density profiles in both CDM+hydro (bottom left panel) and SIDM+hydro simulations (bottom right panel). This improvement can be quantified using using a quality-of-fit function
\begin{equation}
    Q^2 = \frac{1}{N_{\rm bins}}\sum_{i=1}^{N_{\rm bins}}[\ln{\rho(r_i)}-\ln{\rho_{\rm model}(r_i)}]^2
\label{eqn:quality}
\end{equation}
as in \citet{navarro_diversity_2010}; the $Q$ values of the fits for core-Einasto and \abg\ profiles for each simulation are presented in Table~\ref{tab:CDM-SIDM-parameters}. In each case --- both for CDM and SIDM runs with hydrodynamics --- the \abg\ profile fits at least as well as the core-Einasto, and in many cases, it fits significantly better. 

%%%%%%%%%%%%%%%%%%%%%%%%%%%%%%%%%%%%%%%%%%%%%%%%%%
%%%%%%%%%%%%%%%%%%%%%%%%%%%%%%%%%%%%%%%%%%%%%%%%%%

%%%%%%%%%%%%%%%%%%%%%%%%%%%%%%%%%%%%%%%%%%%%%%%%%%
\begin{figure}
    \centering
    \includegraphics[width=\columnwidth]{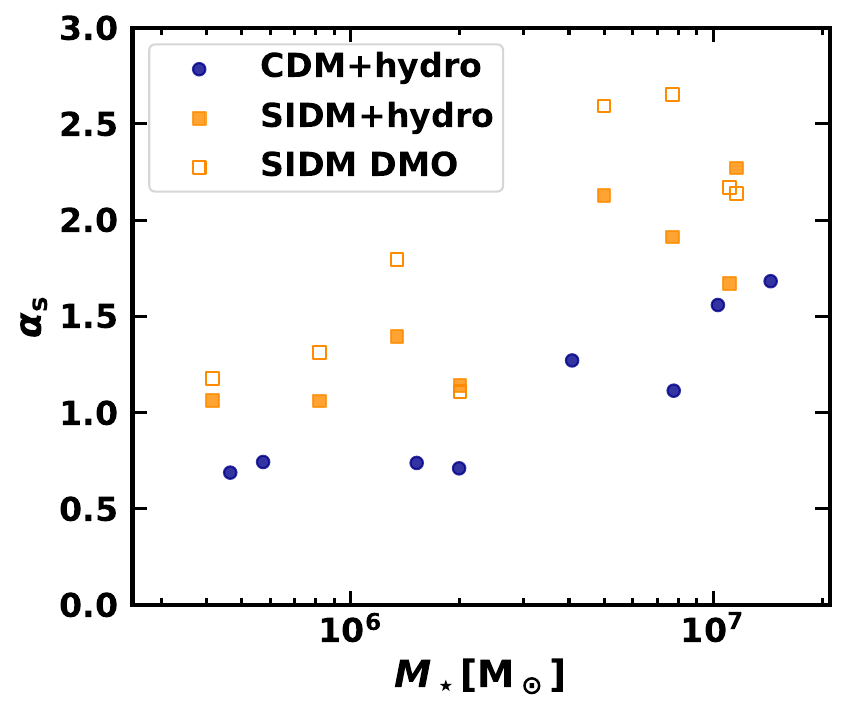}
    \caption{The best-fit value for the shape parameter $\alpha_{\rm s}$ in the \abg\ profile (Eq.~\ref{eqn:alphabetagamma}) as a function of stellar mass $\Mstar$ in CDM (circles) and SIDM (squares). SIDM DMO results are plotted at the stellar mass of the corresponding SIDM+hydro runs. Higher values of $\alpha_{\rm s}$ indicate sharper turnovers, or faster transitions with radius between the slopes of the inner and outer regions of the density profiles. The turnovers are increasingly sharp with higher stellar mass and consistently sharper for SIDM (with or without baryons) than CDM.}
    \label{fig:alphas_vs_Mstar}
\end{figure}
%%%%%%%%%%%%%%%%%%%%%%%%%%%%%%%%%%%%%%%%%%%%%%%%%%

The reason for the improved fits using the \abg\ profile is its inclusion of a shape parameter ($\alpha_{\rm s}$). The form of the core-Einasto profile is such that the transition to a core must occur relatively slowly with decreasing radius. On the other hand, the \abg\ profile can transition rapidly between a steeper outer envelope and a shallow inner core, better fitting the SIDM density profiles. Figure~\ref{fig:alphas_vs_Mstar} shows the best-fit value of $\alpha_{\rm s}$ as a function of stellar mass $\Mstar$. Higher values of $\alpha_{\rm s}$ indicate a faster transition with radius between the slopes of the inner and outer regions of the density profile. The figure demonstrates that the transition speed increases with stellar mass and is consistently higher for SIDM (both with baryons and in dark matter only) than in CDM. Moreover, even in CDM, the value of $\alpha_{\rm s}=1$ (as found in NFW) proves to be too large for the lower stellar mass systems and too small for the higher stellar mass systems. 

The systematic dependence of $\alpha_{\rm s}$ with stellar mass is intuitive for the CDM runs; for the SIDM runs, it is less obvious, as even the SIDM DMO runs show a trend. Rather than a true dependence on $M_{\star}$, it is likely the dependence in the SIDM case comes through a connection to the initial central density: as shown in \citet{fitts_fire_2017} and \citet{fitts_dwarf_2019}, there is a strong correlation between $V_{\rm max}(z=0)$ and $M_\star$ for this sample of simulated galaxies, meaning the halos with higher central densities (and therefore more effective self-interactions) are the ones with sharper density profile changes and higher values of $\alpha_{\rm s}$. Finally, we note that the SIDM DMO runs tend to have slightly larger values of $\alpha_{\rm s}$ than their paired SIDM-hydro versions, indicating a slightly faster transition from inner core to outer cusp. This finding is consistent with the results described above for $r_{-1}$, a further indication that feedback plays a small but non-zero role in establishing the density profiles of SIDM halos and results in a slightly less abrupt transition from the core to the outer profile as compared to SIDM DMO.

\subsection{Dependence on self-interaction cross section}\label{subsec:crosssection}

While our analysis has demonstrated a distinguishing feature in the density profiles of CDM and SIDM dark matter halos, our analysis is limited to eight halos and one cross section, $\sigma/m = 1\ {\rm cm}^2\ {\rm g}^{-1}$. In order to investigate how these results are affected by different dark matter self-interaction cross sections, we re-ran simulations of one halo (m10e, which lies in the middle of the range of stellar masses in our suite) in CDM and in SIDM with $\sigma/m = 0.1, 1,\ \mathrm{and}\ 10 \ {\rm cm}^2\ {\rm g}^{-1}$, hereafter referred to as \sidma,\ \sidm,\ and \sidmb. These simulations were run using the same initial conditions, resolution, and baryonic physics described in \citet{fitts_fire_2017}, and for consistency in comparing the dependence on self-interaction cross section, all mentions of halo m10e in this paper refer to our re-runs of the simulation, rather than the simulation first presented in \citet{fitts_fire_2017}.

\begin{figure*}
    \centering
    \includegraphics[width=\columnwidth]{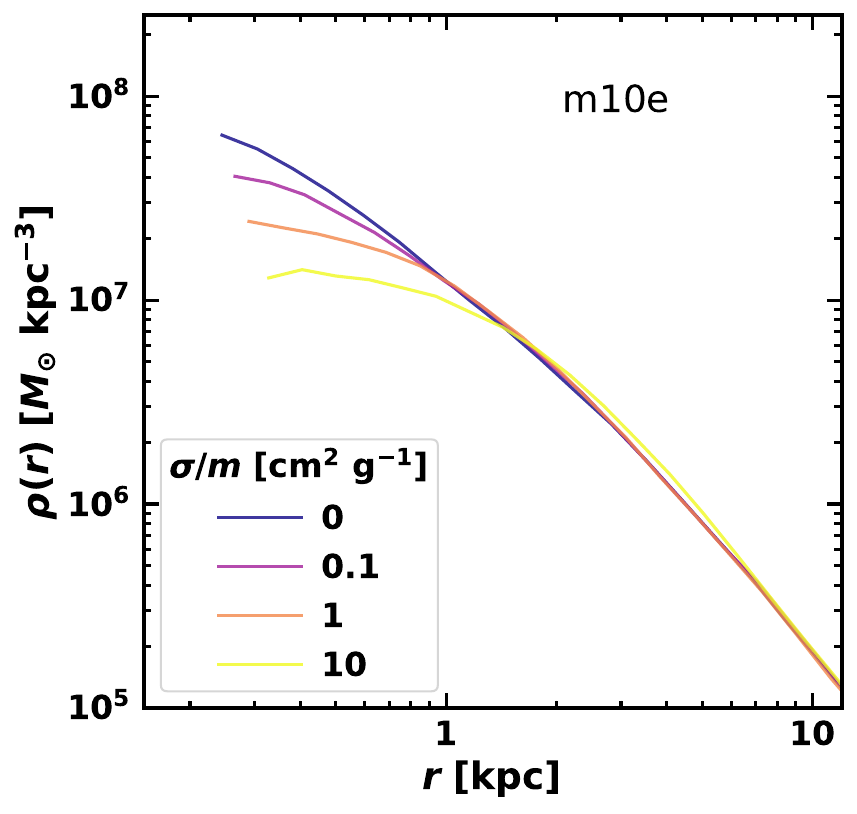}
    \includegraphics[width=\columnwidth]{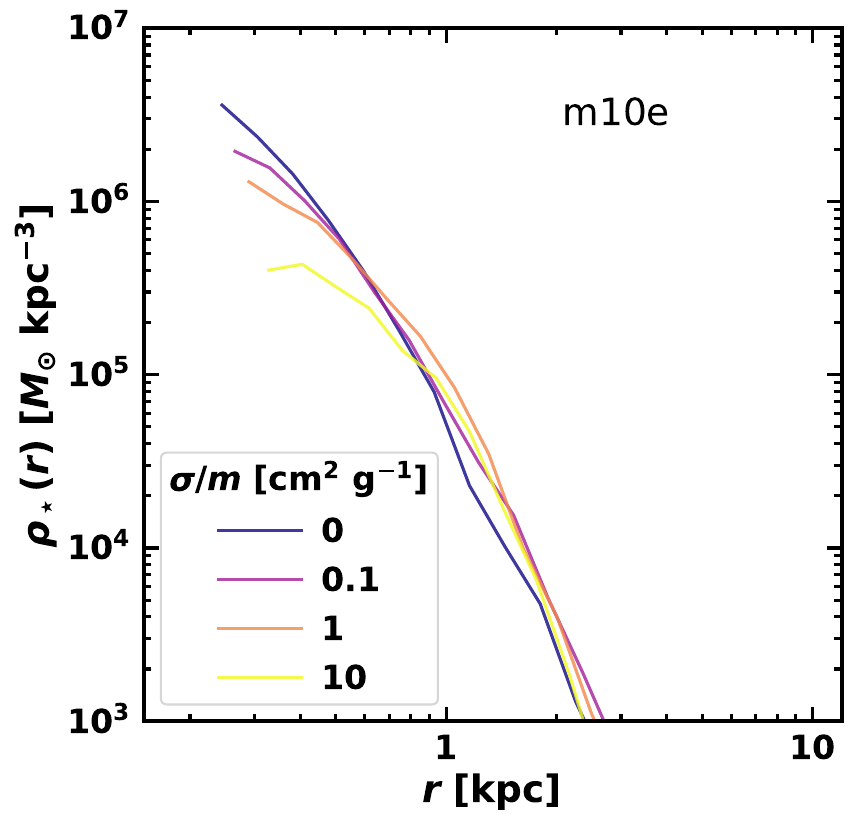}
    \caption{\textit{Left:} Dark matter density profiles of halo m10e with different interaction cross sections indicated by line color. Increasing the interaction cross section decreases the central density. The difference in the central density will also appear in the circular velocity profiles; however, even the two most extreme cross sections lie within error bars of current observations.
    \textit{Right:} The corresponding stellar density profiles show that higher interaction cross sections result in a reduction in the central stellar densities, however this difference would be difficult to distinguish observationally given the scatter in central densities at fixed stellar mass seen in CDM.
    }
    \label{fig:m10e_SIDM}
\end{figure*}

Density profiles for the dark matter distribution and stars for these SIDM runs are plotted in Fig.~\ref{fig:m10e_SIDM}, along with those for the CDM ($\sigma/m=0\ {\rm cm}^2\ {\rm g}^{-1}$) simulation. The features of the \sidma\ and \sidmb\ density profiles are consistent with the trends seen when increasing $\sigma/m$ from 0 to $1\ {\rm cm}^2\ {\rm g}^{-1}$: the central dark matter density is reduced and the core radius is increased with increasing cross section. The value of $r_{-1}$, a proxy for core size, is $1.43\,{\rm kpc}$ in the \sidmb\ run, approximately 1.6 times the value of $0.92\,{\rm kpc}$ from the \sidm\ simulation. The transition between the outer and inner profile is more abrupt as well: $r_{-2}/r_{-1}  \approx 2.9$ for \sidmb\ as compared to $r_{-2}/r_{-1}  \approx 3.8$ for \sidm. We have also explored fitting core-Einasto and \abg\ profiles to the \sidma\ and \sidmb\ runs (with values listed in the bottom section of Table~\ref{tab:CDM-SIDM-parameters}). The results paint a similar picture: the profile transition parameter $\alpha_\mathrm{s}$ is much larger in the \sidmb\ case compared to the \sidm\ version, 1.61 versus 1.14, reinforcing the faster transition from core to outer profile. The \sidma\ simulation has a smaller value of $\alpha_{\rm s}=0.55$ $\alpha_{\rm s}=0.90$. In the stellar density profiles, the most prominent feature is the reduced value of the central stellar density with increasing cross section (right panel of Fig.~\ref{fig:m10e_SIDM}).

In addition to the larger core and sharper turnover in \sidmb\ relative to \sidm, comparing the full baryonic physics version of \sidmb\ to its DMO simulation reveals that the DMO version has slightly larger values of $\alpha_{\rm s}$ (2.1 versus 1.6) and $r_{-1}$ (1.9 versus 1.4 kpc) compared to the simulation with baryons. This is consistent with our finding that, while self-interactions dominate over galaxy formation physics in modifying the density profile relative to CDM DMO expectations, baryonic feedback causes some smoothing in the transition from the core to the outer profile. While a full exploration of cross sections and halo masses is beyond the scope of this work, it would be valuable to map out at what cross sections and halo masses galaxy formation physics dominates over self-interactions, and vice versa, in establishing the properties of density profiles for SIDM halos.

\section{Conclusions}\label{sec:Conclusions}
% The last numbered section should briefly summarise what has been done, and describe the final conclusions which the authors draw from their work.

The distribution of dark matter in dwarf galaxies is not fully understood, especially when it comes to the mismatch between higher central densities predicted by collisionless DMO simulations and the diversity of central dark matter densities implied from observations. This paper investigates simulations of two potential solutions to the central density problem: inclusion of baryonic physics and an alternative model of cold dark matter with self-interactions. We focus on FIRE-2 simulations of dwarf galaxies with $5\times10^{-5} \lesssim \Mstar/\Mhalo \lesssim 10^{-3}$, a stellar mass fraction range in which baryonic feedback does not significantly affect dark matter halo structure, according to simulations that use the FIRE feedback model \citep{chan_impact_2015,onorbe_how_2014,robles_sidm_2017,lazar_dark_2020} and other models \citep{di_cintio_mass-dependent_2014, tollet_nihao_2016}. 

At this stellar mass range, SIDM halos with the self-interaction cross section that is the primary focus of our study---$\sigma/m = 1\ {\rm cm}^2\ {\rm g}^{-1}$---have lower central densities and are clearly distinguishable from their CDM counterparts. The logslopes of the density profiles show that in comparison to CDM, SIDM profiles transition more quickly with increasing radius from a shallow central core to a steep outer profile. We find that this rapid transition with radius in the slope of the density profile for SIDM is better modeled by a three-parameter $\alpha\beta\gamma$ fit than by the three-parameter core-Einasto fit designed for halo profiles with feedback-induced cores. A shape parameter $\alpha_{\rm s}$ in the \abg\ profile provides a better fit for SIDM profiles, and the higher best-fit values of $\alpha_{\rm s}$ for SIDM also show a sharper turnover in SIDM than in CDM density profiles.

While this paper focuses on self-interaction cross sections of $\sigma/m = 1\ {\rm cm}^2\ {\rm g}^{-1}$, the behavior noted here---lower central densities and faster transitions from shallow inner profiles to steep outer profiles compared to CDM---is expected to scale with the interaction rate $\Gamma=\rho_{\rm dm} \,v\,\sigma/m$. Indeed, re-simulating an individual halo (m10e) with hydrodynamics and a self-interaction cross section of $\sigma/m = 10\ {\rm cm}^2\ {\rm g}^{-1}$ revealed a larger core and a sharper transition than the lower interaction cross section case. We expect this trend to continue with increasing cross section values until core collapse commences \citep{elbert_core_2015}. This scaling also gives an intuitive understanding of the faster inner to outer profile transition we note in SIDM halos. Self-interaction rates depend on the local dark matter density, and the probability of interaction (and therefore of affecting the halo profile) changes rapidly with radius for SIDM halos. By contrast, feedback-induced cores seen in CDM halos rely on rapid potential fluctuations that drive gas motions, a process that is less sensitive to the local dark matter density.

The differing predictions indicate the possibility of distinguishing between these two models of dark matter in this stellar mass range ($5\times10^{-5} \lesssim \Mstar/\Mhalo \lesssim 10^{-3}$) even subject to the baryonic physics operating in these galaxies. 
Due to complex dynamics and limited data, accurately modeling the dark matter distribution of dwarf galaxies from observations remains a challenge; constraining whether the center of a halo is cored or cusped is especially difficult with data sets containing fewer than 10,000 stars \citep{chang_dark_2021}. However, future surveys \citep{takada_extragalactic_2014,bundy_fobos_2019,MSE_detailed_2019} and improved modeling (e.g. \citealt{nguyen_uncovering_2023}) may soon enable distinguishing between observations of dwarf galaxy halos with central cusps and cores. With precise observations of dwarf galaxies, the results presented in this paper will help determine whether an inferred dark matter density profile is better explained by CDM or SIDM, thereby providing an important clue to the nature of dark matter.

\section*{Acknowledgements}
We thank the referee for insightful comments that helped improved this paper. 
We thank Manoj Kaplinghat for helpful discussions. MCS acknowledges support from the National Science Foundation Graduate Research Fellowship Program under Grant No. DGE 2137420 and from the UT Austin Astronomy Department REU Program "Frontier Research and Training in Astronomy for the 21st Century" funded by NSF grant AST 1757983 from the NSF REU program and the Department of Defense ASSURE program. MBK acknowledges support from NSF CAREER award AST-1752913, NSF grants AST-1910346 and AST-2108962, NASA grant 80NSSC22K0827, HST-GO-16686, HST-AR-17028, HST-AR-17043, JWST-GO-03788, aand JWST-AR-06278 from the Space Telescope Science Institute, which is operated by AURA, Inc., under NASA contract NAS5-26555; and from the Samuel T. and Fern Yanagisawa Regents Professorship in Astronomy at UT Austin. XS acknowledges the support from NASA grant JWST-AR04814. LN is supported by the Sloan Fellowship, the NSF CAREER award 2337864, NSF award 2307788, and by the NSF award PHY2019786 (The NSF AI Institute for Artificial Intelligence and Fundamental Interactions, http://iaifi.org/).
We thank the developers of the Python packages used in preparing this paper: \textsc{NumPy} \citep{harris_array_2020}, \textsc{SciPy} \citep{virtanen_scipy_2020}, and \textsc{matplotlib} \citep{hunter_matplotlib_2007}. This work used computational resources of the University of Texas at Austin and the Texas Advanced Computing Center (TACC; http://www.tacc.utexas.edu), the NASA Advanced Supercomputing (NAS) Division and the NASA Center for Climate Simulation (NCCS), and the Extreme Science and Engineering Discovery Environment (XSEDE), which is supported by National Science Foundation grant number OCI-1053575.

%%%%%%%%%%%%%%%%%%%%%%%%%%%%%%%%%%%%%%%%%%%%%%%%%%
\section*{Data Availability}

The FIRE-2 simulations are publicly available \citep{wetzel_public_2023} at \url{http://flathub.flatironinstitute.org/fire}. Additional FIRE simulation data is available at \url{https://fire.northwestern.edu/data}. A public version of the \textsc{gizmo} code is available at \url{http://www.tapir.caltech.edu/~phopkins/Site/GIZMO.html}. Data products from this paper will be made available upon reasonable request to the corresponding author.


\bibliographystyle{mnras}
\bibliography{references}

%%%%%%%%%%%%%%%%%%%%%%%%%%%%%%%%%%%%%%%%%%%%%%%%%%

%%%%%%%%%%%%%%%%% APPENDICES %%%%%%%%%%%%%%%%%%%%%

\appendix
\section{Profile fits}
\label{sec:appendix}

Table~\ref{tab:CDM-SIDM-parameters} contains the stellar masses, defined as $\Mstar(<0.1R_{\rm vir})$, and best-fit parameters for the core-Einasto (Eq.~\ref{eqn:coredEinasto}) and \abg~(Eq.~\ref{eqn:alphabetagamma}) analytical density profiles for the CDM and SIDM halo density profiles at redshift $z=0$. Further properties of these halos at $z=0$ can be found in \citet{fitts_fire_2017}.

\begin{table*}
    \caption{Best-fit parameters for the core-Einasto (Eq.~\ref{eqn:coredEinasto}) and $\alpha\beta\gamma$ (Eq.~\ref{eqn:alphabetagamma}) profile fits for CDM+hydro, SIDM+hydro ($\sigma/m = 1\ {\rm cm^2\ g^{-1}}$) halos, and the m10e halos with two additional self-interaction cross-sections (noted in the halo name column). Setting the shape parameter $\hat{\alpha}=0.16$ in Eq.~\ref{eqn:coredEinasto} results in a three parameter fit where where $r_{\rm c}$ is the dark matter core radius and $\tilde{r}_{\rm s}$ and $\tilde{\rho}_{\rm s}$ are radius and density free parameters. $r_{\rm c}$ values less than $\sim0.2\ {\rm kpc}$ indicate that the fitted core radius is smaller than the convergence radius. In the $\alpha\beta\gamma$ profile (Eq.~\ref{eqn:alphabetagamma}), setting $\beta_{\rm s}=2.5$ and $\gamma_{\rm s} = 0$ results in a three-parameter fit where $\alpha_{\rm s}$ describes the rate at which the slope changes with radius between the inner and outer regions of the halo, and $r_{\rm s}$ and $\rho_{\rm s}$ are the scale radius and scale density. The quality-of-fit $Q$ is given by Eq.~\ref{eqn:quality}.
    }
    
    \begin{tabular*}{\textwidth}{@{\extracolsep{\fill}}*{10}{r}@{}}
        \hline
        
        & & \multicolumn{4}{c}{core-Einasto} & \multicolumn{4}{c}{$\alpha\beta\gamma$}\\
        \cmidrule(lr){3-6}\cmidrule(lr){7-10}
        
        CDM halo & $\Mstar [\Msun]$ & $r_{\rm c} [\rm kpc]$ & $\tilde{r}_{\rm s} [\rm kpc]$ & $\tilde{\rho}_{\rm s} [\Msun\ {\rm kpc}^{-3}]$ & $Q_{\rm cEin}$
        & $\alpha_{\rm s}$ & $r_{\rm s} [\rm kpc]$ & $\rho_{\rm s} [\Msun\ {\rm kpc}^{-3}]$ & $Q_{\alpha\beta\gamma}$\\[2pt]

        \hline
        m10b
&
4.7$\ \times\ 10^{5}$
&
0.00
&
4.21
&
1.3$\ \times\ 10^{6}$
&
0.107
&
0.69
&
0.37
&
9.2$\ \times\ 10^{8}$
&
0.070
 \\[2pt]
        m10c
&
5.7$\ \times\ 10^{5}$
&
0.03
&
5.07
&
7.9$\ \times\ 10^{5}$
&
0.086
&
0.74
&
0.57
&
3.2$\ \times\ 10^{8}$
&
0.082
 \\[2pt]
        m10d
&
1.5$\ \times\ 10^{6}$
&
0.00
&
3.83
&
1.5$\ \times\ 10^{6}$
&
0.078
&
0.74
&
0.38
&
7.8$\ \times\ 10^{8}$
&
0.083
 \\[2pt]
        m10e
&
2.0$\ \times\ 10^{6}$
&
0.05
&
6.01
&
5.8$\ \times\ 10^{5}$
&
0.069
&
0.71
&
0.66
&
2.7$\ \times\ 10^{8}$
&
0.043
 \\[2pt]
        m10f
&
4.1$\ \times\ 10^{6}$
&
0.14
&
2.46
&
4.9$\ \times\ 10^{6}$
&
0.112
&
1.27
&
0.64
&
1.8$\ \times\ 10^{8}$
&
0.129
 \\[2pt]
        m10h
&
7.8$\ \times\ 10^{6}$
&
0.18
&
3.51
&
2.8$\ \times\ 10^{6}$
&
0.101
&
1.11
&
0.79
&
1.5$\ \times\ 10^{8}$
&
0.040
 \\[2pt]
        m10k
&
1.0$\ \times\ 10^{7}$
&
0.65
&
2.30
&
8.8$\ \times\ 10^{6}$
&
0.087
&
1.56
&
1.20
&
4.8$\ \times\ 10^{7}$
&
0.090
 \\[2pt]
        m10m
&
1.4$\ \times\ 10^{7}$
&
0.61
&
2.00
&
1.3$\ \times\ 10^{7}$
&
0.102
&
1.68
&
1.09
&
5.9$\ \times\ 10^{7}$
&
0.130
 \\[2pt]
        \hline
        SIDM halo \\
        \hline
        m10b
&
4.2$\ \times\ 10^{5}$
&
0.12
&
3.75
&
1.8$\ \times\ 10^{6}$
&
0.144
&
1.06
&
0.75
&
1.3$\ \times\ 10^{8}$
&
0.095
 \\[2pt]
        m10c
&
8.2$\ \times\ 10^{5}$
&
0.23
&
4.05
&
1.4$\ \times\ 10^{6}$
&
0.120
&
1.06
&
0.94
&
7.8$\ \times\ 10^{7}$
&
0.058
 \\[2pt]
        m10d
&
1.3$\ \times\ 10^{6}$
&
0.33
&
2.54
&
4.6$\ \times\ 10^{6}$
&
0.119
&
1.40
&
0.92
&
7.0$\ \times\ 10^{7}$
&
0.105
 \\[2pt]
        m10e
&
2.0$\ \times\ 10^{6}$
&
0.44
&
4.49
&
1.3$\ \times\ 10^{6}$
&
0.133
&
1.14
&
1.31
&
3.8$\ \times\ 10^{7}$
&
0.061
 \\[2pt]
        m10f
&
5.0$\ \times\ 10^{6}$
&
0.39
&
1.91
&
1.1$\ \times\ 10^{7}$
&
0.196
&
2.13
&
0.94
&
6.3$\ \times\ 10^{7}$
&
0.101
 \\[2pt]
        m10h
&
1.1$\ \times\ 10^{7}$
&
0.58
&
2.52
&
7.4$\ \times\ 10^{6}$
&
0.149
&
1.67
&
1.21
&
4.9$\ \times\ 10^{7}$
&
0.039
 \\[2pt]
        m10k
&
1.2$\ \times\ 10^{7}$
&
1.24
&
1.55
&
3.1$\ \times\ 10^{7}$
&
0.146
&
2.27
&
1.54
&
2.5$\ \times\ 10^{7}$
&
0.086
 \\[2pt]
        m10m
&
7.7$\ \times\ 10^{6}$
&
0.82
&
1.82
&
1.8$\ \times\ 10^{7}$
&
0.123
&
1.91
&
1.27
&
4.1$\ \times\ 10^{7}$
&
0.119
 \\[2pt]
        \hline
        m10e\_0.1
&
2.0$\ \times\ 10^{6}$
&
0.20
&
5.18
&
8.6$\ \times\ 10^{5}$
&
0.085
&
0.90
&
0.95
&
9.2$\ \times\ 10^{7}$
&
0.033
 \\[2pt]
        m10e\_10
&
1.3$\ \times\ 10^{6}$
&
1.03
&
3.14
&
3.7$\ \times\ 10^{6}$
&
0.162
&
1.61
&
1.81
&
1.6$\ \times\ 10^{7}$
&
0.052
 \\[2pt]
    \label{tab:CDM-SIDM-parameters}
    \end{tabular*}    
\end{table*}

% Don't change these lines
\bsp	% typesetting comment
\label{lastpage}
\end{document}